\documentclass[showpacs,aps,prb,reprint,superscriptaddress,preprintnumbers]{revtex4-2}

\usepackage{amsmath}
\usepackage{amssymb}
\usepackage{graphicx}
\usepackage{dcolumn}
\usepackage[dvipsnames]{xcolor}
\usepackage{amsthm,amssymb}
\usepackage{mathrsfs}
\usepackage{color}
\usepackage{soul}
\usepackage{endnotes}
\usepackage{bm}
\usepackage{epsfig}
\usepackage{array}
\usepackage{ulem}
\usepackage{lineno}

\newcommand{\CsVSb}{{CsV$_{3}$Sb$_{5}$}}
\newcommand{\RbVSb}{{RbV$_{3}$Sb$_{5}$}}
\newcommand{\KVSb}{{KV$_{3}$Sb$_{5}$}}
\newcommand{\AVSb}{{AV$_{3}$Sb$_{5}$}} 
\newcommand{\parallelsum}{\mathbin{\!/\mkern-5mu/\!}}

\begin{document}


\title{Similarities and Differences in the Fermiology of Kagome Metals\\ AV$_{3}$Sb$_{5}$ (A=K, Rb, Cs) Revealed by Shubnikov-de Haas Oscillations}

\author{Zheyu~Wang$^\S$}
\author{Wei~Zhang$^\S$}
\author{Lingfei~Wang}
\author{Tsz~Fung~Poon}
\author{Chun~Wai~Tsang}
\author{Wenyan~Wang}
\author{Jianyu Xie}
\affiliation{Department of Physics, The Chinese University of Hong Kong, Shatin, Hong Kong, China}
\author{Xuefeng~Zhou}
\author{Yusheng~Zhao}
\author{Shanmin~Wang}
\affiliation{Department of Physics, Southern University of Science and Technology, Shenzhen, Guangdong, China}
\author{Ming-zhong~Ai}
\affiliation{Department of Physics, The Chinese University of Hong Kong, Shatin, Hong Kong, China}
\author{Kwing~To~Lai}
\affiliation{Department of Physics, The Chinese University of Hong Kong, Shatin, Hong Kong, China} 
\affiliation{Shenzhen Research Institute, The Chinese University of Hong Kong, Shatin, Hong Kong, China}
\author{Swee~K.~Goh}
\email[]{skgoh@cuhk.edu.hk}
\affiliation{Department of Physics, The Chinese University of Hong Kong, Shatin, Hong Kong, China}



\date{\today}

\begin{abstract}

Materials with AV$_3$Sb$_5$ (A=K, Rb, Cs) stoichiometry are recently discovered kagome superconductors  with the electronic structure featuring a Dirac band, van Hove singularities and flat bands. These systems undergo anomalous charge-density-wave (CDW) transitions at $T_{\rm CDW}\sim80-100$~K, resulting in the reconstruction of the Fermi surface from the pristine phase. Although comprehensive investigations of the electronic structure via quantum oscillations (QOs) have been performed on the sister compounds CsV$_3$Sb$_5$ and RbV$_3$Sb$_5$, a detailed QO study of KV$_3$Sb$_5$ is so far absent. Here, we report the Shubnikov-de Haas QO study in KV$_3$Sb$_5$. We resolve a large number of new frequencies with the highest frequency of 2202 T (occupying $\sim$54\% of the Brillouin
zone area in the $k_x$-$k_y$ plane). The Lifshitz-Kosevich analysis further gives relatively small cyclotron effective masses, and the angular dependence study reveals the two-dimensional nature of the frequencies with a sufficient signal-to-noise ratio. Finally, we compare the QO spectra for all three AV$_3$Sb$_5$ compounds collected under the same conditions, enabling us to point out the similarities and differences across these systems. Our results fill in the gap of the QO study in KV$_3$Sb$_5$ and provide valuable data to understand the band structure of all three members of AV$_3$Sb$_5$.

\end{abstract}

\maketitle

\section{Introduction} 
The kagome metals \AVSb\ (A=K, Rb, Cs) with vanadium atoms arranged in a perfect kagome net configuration host flat electronic bands, van Hove singularities and Dirac points~\cite{Ortiz2019,Neupert2021,Li2021,Liu2021,Kang2022a}. As such, these systems offer an ideal platform to explore the interplay between the band topology and electronic correlation~\cite{Kiesel2012,Kiesel2013,Wang2013}. Interestingly, these metals enter anomalous charge-density-wave (CDW) phases ($T_{\rm CDW}\sim80-100$~K) and exhibit superconductivity ($T_{\rm c}\sim0.9 - 2.7$~K) on cooling~\cite{Li2021,Liang2021,Zhao2021,Liu2021,Hu2021,Uykur2021,Zhou2021,Jiang2021,Kang2022a,Wu2022,Lou2022,Kato2023}. The CDW order and superconductivity exhibit an interesting interplay, as revealed by high-pressure experiments~\cite{Yu2021,Du2021,Chen2021a,Wang2021a}. Besides, recent studies detected the time-reversal symmetry breaking and electronic nematicity in the CDW phase, evoking extensive exploration of the possible unconventional superconductivity (SC) in \AVSb~\cite{Yu2021c,Hu2022,Mielke2022,Khasanov2022,Nie2022,Li2022,Xu2022}.

Although all three members of \AVSb\ take the same crystal structure in the pristine phase, both the superconducting and the CDW states exhibit crucial differences. Apart from the obvious difference in $T_c$, the superconducting gap and its behaviour against pressure are different. For \KVSb\ and \RbVSb, the superconducting gap has been reported to be nodal at ambient pressure, but as pressure increases the gap becomes nodeless~\cite{Guguchia2023}. For \CsVSb, accumulating evidence points to the nodeless gap at ambient pressure, and the application of pressure does not affect the nodeless nature of the superconducting gap~\cite{Gupta2022,Gupta2022a,Duan2021,Shan2022,Roppongi2023,Mu2021,Zhang2023}. Concerning the CDW state, important differences have also been noted. In the CDW state of \CsVSb, a complicated superlattice involving the stacking of “Star of David” (SoD) and  trihexagonal (TrH) in-plane distortions has been proposed, quadrupling the periodicity along the $c$-axis (such a distortion is denoted as ``2$\times$2$\times$4")~\cite{Ortiz2021, Kang2022a}. On the other hand, the charge order in \KVSb\ and \RbVSb\ adopt a staggered TrH configuration with a $\pi$ shift between the adjacent layers, resulting in a 2$\times$2$\times$2 distortion~\cite{Kang2022a}. Given that the SC phase arises from the CDW state, as $T_c<T_{\rm CDW}$ for all three systems, it is important to understand the normal state in the presence of the CDW for an eventual understanding of the superconductivity.

The CDW order unavoidably plays an important role on the Fermiology of these kagome metals. From the scanning tunnelling microscopy (STM) and angle-resolved photoemission spectroscopy (ARPES) measurements, an energy gap opens near the Fermi level during the CDW transition and the Fermi surfaces are reconstructed~\cite{Liang2021,Xu2021,Luo2022,Hu2022b,Kang2022a}. Due to the complicated CDW order, the Fermi surface in the CDW phase can be very different from the pristine phase. Quantum oscillations (QOs) have been most heavily studied in the CDW phase of \CsVSb, revealing rich QO spectra with frequencies up to 9930~T \cite{Chapai2022,Ortiz2021,Fu2021,Yu2021,Yu2021b,Shrestha2022,Huang2022,Chen2022,Gan2021}.
Surprisingly, a significantly simpler QO spectrum was reported in \KVSb\ -- only two low frequencies (34.6 T and 148.9 T) were revealed by the Shubnikov-de Haas (SdH) effect in the pioneering work of \KVSb~\cite{Yang2020}. Interestingly, a simple QO spectrum with only two small frequencies was also initially reported in \RbVSb, but a later study by some of us uncovered a much richer spectrum with the largest maximum QO frequency extending beyond 2000~T~\cite{Wang2022a,Shrestha2023}. Therefore, it is a pressing issue to revisit the Fermiology of \KVSb\ via QOs.

In this work, we measure the SdH quantum oscillations using high-quality single crystals of \KVSb. We resolve a large number of new frequencies with the highest frequency of 2202 T. This frequency is significant, as it represents $\sim$54\% of the reconstructed Brillouin zone area in the $k_x$-$k_y$ plane. The Lifshitz-Kosevich analysis further gives relatively small cyclotron effective masses, and the angular dependence study reveals the two-dimensional nature for the frequencies with a sufficient signal-to-noise ratio. Our results show that the QO spectra in all three compounds are complicated, likely resulting from the complex CDW order, and that large Fermi surfaces beyond the expectation from the initial QO reports exist in these metals.


\section{Methods}

Single crystals of \KVSb\ were synthesized from K (ingot, 99.97 $\%$), V (powder, 99.9 $\%$), and Sb (shot, 99.9999 $\%$), using self-flux method similar to Ref.~\cite{Ortiz2019}. Raw materials with the molar ratio of K:V:Sb = 5:3:14 were sealed in a stainless steel jacket in an argon-filled glovebox and then moved to the furnace for heat treatment. The mixture was first heated to 1000~$^{\circ}$C at the rate of 20~$^{\circ}$C/h. After being held at 1000 $^{\circ}$C for 24 h, it was cooled to 900 $^{\circ}$C at 50 $^{\circ}$C/h, followed by a further cooling to 400 $^{\circ}$C at 2 $^{\circ}$C/h. The as-grown single crystals were millimeter-sized shiny plates. X-ray diffraction measurements  were performed at room temperature by using a Rigaku X-ray diffractometer with Cu$K_\alpha$ radiation. The chemical compositions were characterized by a JEOL JSM-7800F scanning electron microscope equipped with an Oxford energy-dispersive X-ray (EDX) spectrometer.

\begin{figure}[!t]
      \resizebox{8.8cm}{!}{
              \includegraphics{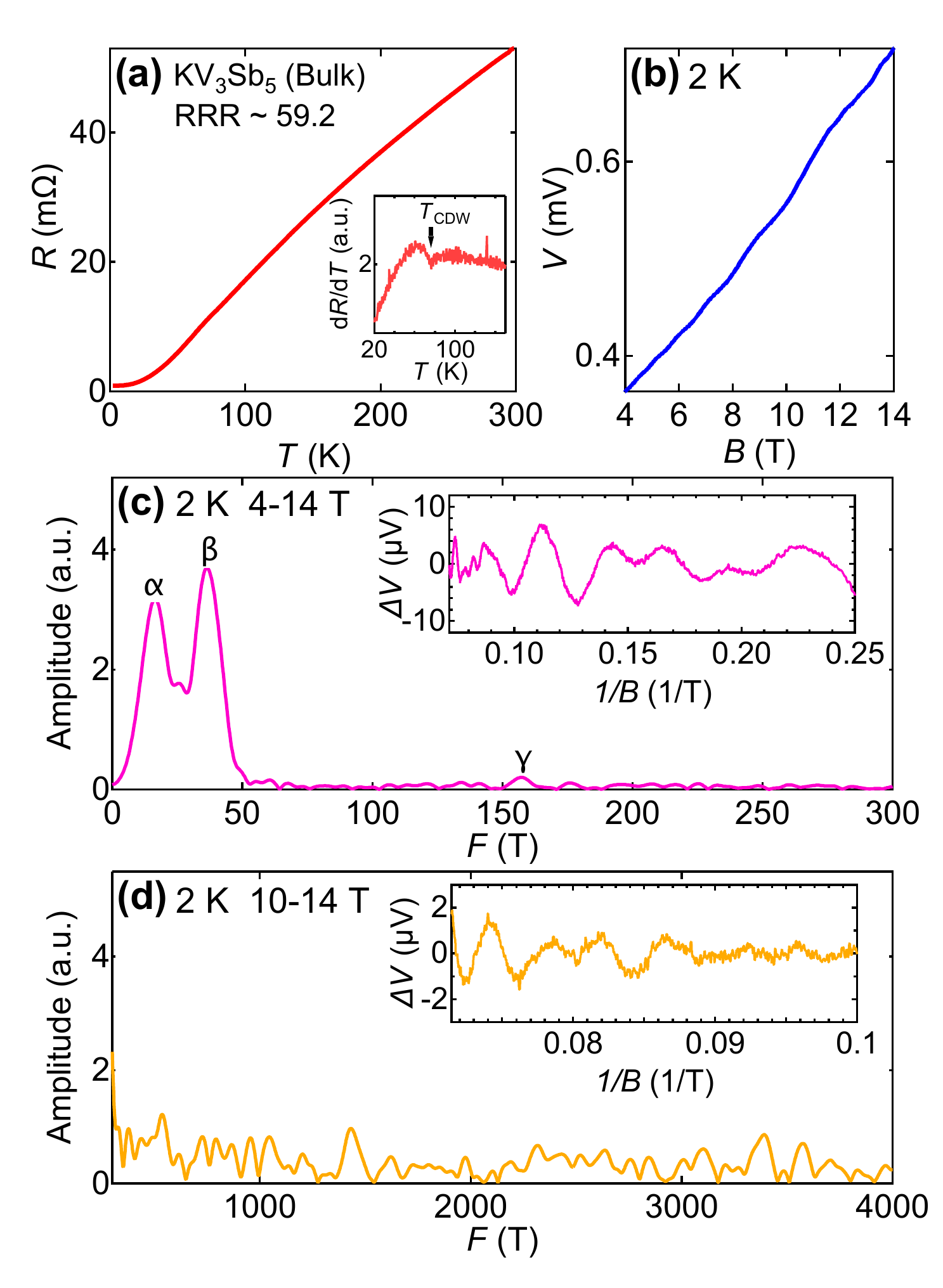}}                				
              \caption{\label{BulkQO}  (a) Temperature dependence of resistance for a bulk \KVSb\ sample. The RRR is 59.2. The inset is the temperature dependence of d$R$/d$T$, showing the determination of $T_{\rm{CDW}}$. (b) Raw data of the lock-in voltage against the magnetic field at 2~K with $B\,\parallelsum \,c$.  FFT spectra of the bulk sample for the oscillation data from (c) 4~T to 14~T and (d) 10~T to 14~T. The insets show the oscillatory signals after removing the background.}
\end{figure}


Magnetotransport measurements were conducted in a Physical Property Measurement System by Quantum Design. For bulk crystals, the electrical contacts were prepared with gold wires and silver paste while for thin flakes, the samples were first exfoliated from bulk crystals and then transferred onto a diamond substrate pre-patterned with conducting electrodes~\cite{Xie2021, Ku2022, Zhang2022, Wang2022a}. The thin flake sample was encapsulated by a h-BN film to avoid oxidization. All operations above were conducted in an argon-filled glovebox. The total thickness of h-BN encapsulated flake was determined by a Bruker BioScope Resolve atomic force microscope (AFM). The thickness of h-BN is determined separately. The flake thickness is then the difference between the two AFM measurements. A Stanford Research 830 lock-in amplifier was used for Shubnikov-de Haas effect measurements. The rotator insert option by Quantum Design was used to tilt the angle between magnetic field and the sample. During the rotation, the current direction is kept perpendicular to the magnetic field.

\section{Results}
\begin{figure*}
    \centering
      \resizebox{15cm}{!}{
              \includegraphics{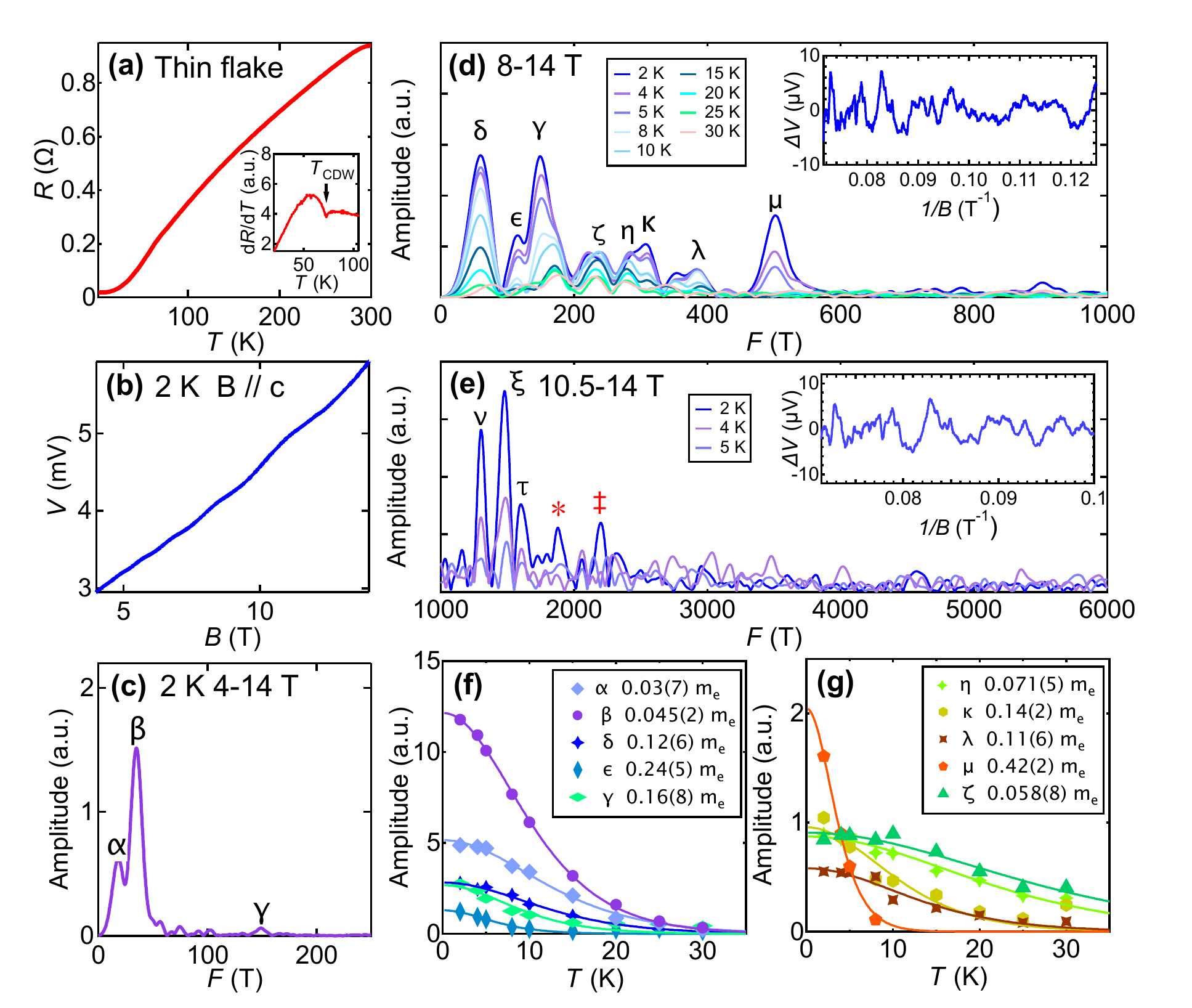}}        \caption{\label{ThinflakeQO}  
          (a) Temperature dependence of the resistance for a thin flake of \KVSb. The inset is the temperature dependence of d$R$/d$T$, allowing the determination of $T_{\rm{CDW}}$. (b) Resistive voltage against the magnetic field at 2 K with $B\,\parallelsum \,c$. FFT spectra of the thin flake for the oscillation data from (c) 4~T to 14~T, (d) 8~T to 14~T and (e) 10.5~T to 14~T. For the peaks labelled by ``$\ast$" and ``$\ddag$", we only observed them at 2~K. The insets in (d) and (e) show the oscillation signals with the background subtracted. (f), (g) Temperature dependence of oscillation amplitudes. The symbols are experimental data and the solid curves are the Lifshitz-Kosevich fits.
              }
\end{figure*}
Figure~\ref{BulkQO}(a) shows the temperature dependence of resistance, $R(T)$, in a bulk \KVSb\ sample. The residual resistivity ratio (RRR, defined as $\rho(300~\rm{K})$/$\rho(2~\rm{K})$) is 59.2, which is the highest among the reported values, indicating the high quality of our sample. On cooling, there is a weak anomaly in $R(T)$ and correspondingly, a dip feature appears in the d$R$/d$T$ curve at around 78 K (inset of Fig.~\ref{BulkQO}(a)), which is consistent with the reported CDW transition~\cite{Du2021,Luo2022,Yang2020}.  

The high purity of the sample enables us to study the SdH effect in \KVSb. As shown in Fig.~\ref{BulkQO}(b), the SdH quantum oscillation signals can be recognized in the field-dependent resistive signals even without removing the background. We further perform the fast Fourier transformation (FFT) analysis. As displayed in Fig.~\ref{BulkQO}(c), apart from $\beta$ (35 T) and $\gamma$ (157 T), which have been revealed by previous study~\cite{Yang2020}, we resolve another new QO frequency ($F_\alpha$=16 T). When a shorter field range (between 10~T and 14~T) is used, which would preferentially enhance the spectral intensities of relatively high QO frequencies, we do not observe conclusive peaks in the FFT spectrum (Fig.~\ref{BulkQO}(d)).

Previous QO studies on the sister compounds \RbVSb\ and \CsVSb\ have uncovered much higher QO frequencies. To search for higher frequencies in \AVSb, it is necessary to improve the signal-to-noise ratio. One avenue would be to extend the maximum field of the experiment, as was adopted by the authors of Refs.~\cite{Shrestha2022,Chapai2022,Huang2022} on \CsVSb.  Here, we choose to enhance the resistive signal strength by optimizing the sample geometry -- we prepare a device using a 80-nm-thick flake of \KVSb\ exfoliated from the bulk crystal. Figures~\ref{ThinflakeQO}(a) and \ref{ThinflakeQO}(b) show the temperature- and field-dependence of the resistive signals in the thin flake of \KVSb, respectively. The magnetic field is applied along the sample $c$-axis. Analyzing the signals carefully, we uncover a rich QO spectrum. As displayed in Fig.~\ref{ThinflakeQO}(c), $F_\alpha$ (18~T), $F_\beta$ (35~T) and $F_\gamma$ (148~T) are resolved using a long field range of $4-14$~T, consistent with the results of the bulk sample ({\it c.f.} Fig.~\ref{BulkQO}(c)). However, new frequencies can be readily observed when narrower field windows are employed, as displayed in Figs.~\ref{ThinflakeQO}(d) and (e). Figure~\ref{ThinflakeQO}(d) shows the FFT spectrum with the field range of $8-14$~T and we resolve $F_\delta= 60$~T, $F_\epsilon= 115$~T, $F_\zeta= 237$~T, $F_\eta= 285$~T, $F_\kappa= 308$~T, $F_\lambda= 385$~T and $F_\mu= 502$~T below 1000~T. In Fig.~\ref{ThinflakeQO}(e) in which a field window of $10.5-14$~T is used, much higher frequencies can be seen: $F_\nu=1304$~T , $F_\xi=1480$~T, $F_\tau=1601$~T and probably $F_\ast=1879$~T, and $F_\ddag$ = 2202~T. For $F_\ast$ and $F_\ddag$, the oscillatory signals already fade away at 4~K, rendering their presence more uncertain.

We next measure QOs at different temperatures and we carefully trace the temperature dependence of the QO amplitudes for ten frequencies, enabling the determination of cyclotron effective masses. As displayed in Figs.~\ref{ThinflakeQO}(f) and \ref{ThinflakeQO}(g), the cyclotron effective masses obtained by the Lifshitz-Kosevich (LK) analysis are small, with the values $\sim$0.1 $m_e$ or below, except for $m_{\epsilon}$ = $0.24~m_e$ and $m_{\mu}$ = $0.42~m_e$. The small effective masses indicate that these frequencies originate from electronic bands with quasilinear nature. Unfortunately, we are unable to determine the effective masses for the peaks larger than 1300~T, as the QO amplitudes decrease rapidly at higher temperatures (see Fig.~\ref{ThinflakeQO}(e)). Based on the effective masses we are able to extract, the rather low effective masses are consistent with the sister kagome compound \CsVSb\ and \RbVSb~\cite{Ortiz2021,Zhang2022,Gan2021,Wang2022a}.

\begin{figure}
    \centering
      \resizebox{8.8cm}{!}{
              \includegraphics[width=1\textwidth]{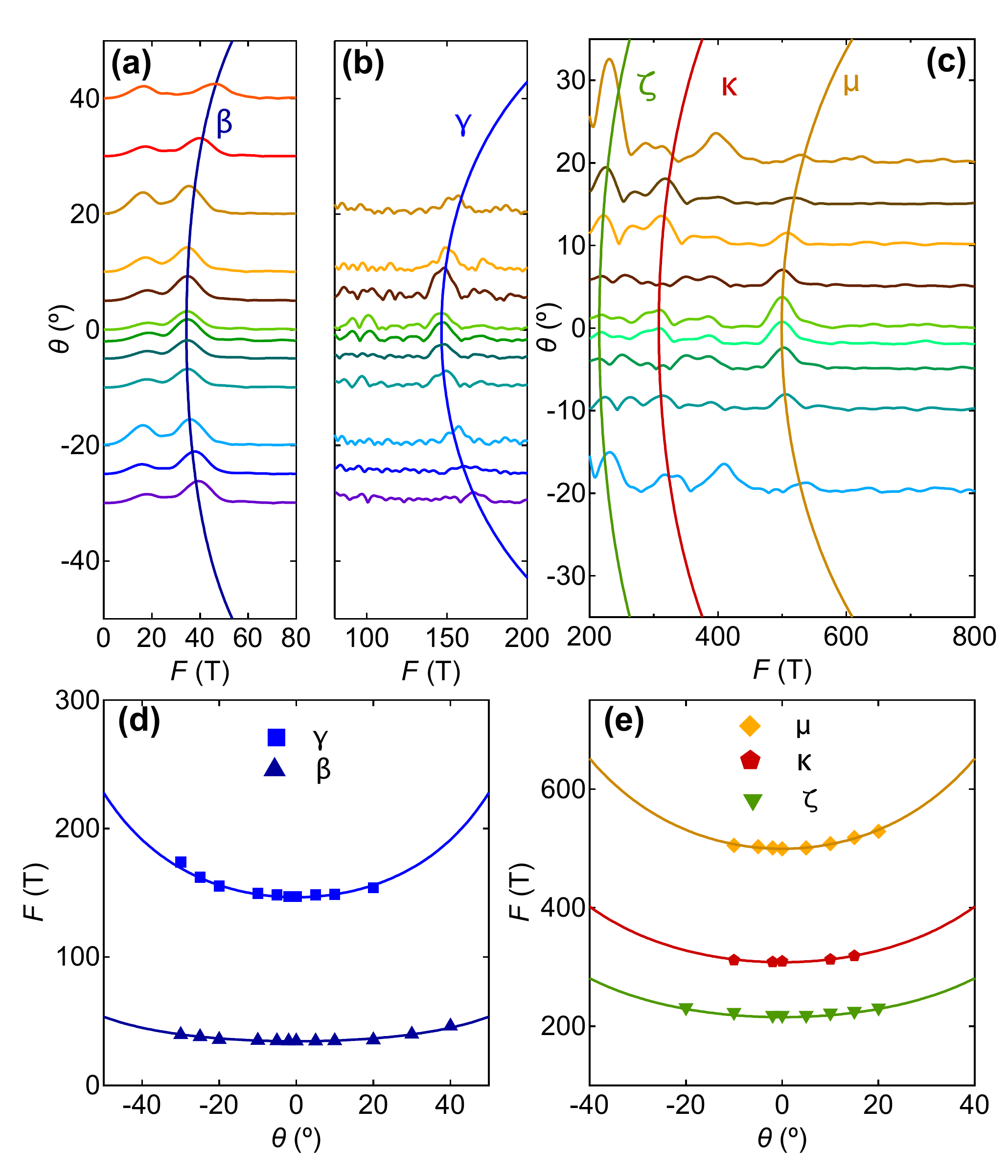}}                
              \caption{\label{Ang_dep} FFT spectra at different field angles for the frequency range of (a) 0 to 80~T, (b) 80 to 200T, (c) 200 to 800~T. (d, e) Angular dependence of QO frequencies. The solid curves are fits using $F(\theta)=F(0^{\circ})/ \cos(\theta)$.  }
\end{figure}

 We now explore the dimensionality of the Fermi surfaces by measuring SdH oscillations at various magnetic field angles. Here $\theta=0$ refers to $B\,\parallelsum \,c$ and the rotation is towards the $ab$ plane. Due to the difference in the amplitudes, FFT spectra are magnified by different factors, and they are divided into three frequency ranges for visual clarity, see Figs.~\ref{Ang_dep}(a) -- \ref{Ang_dep}(c). When $\theta$ increases, all frequencies shift to higher values and the intensities gradually decrease. Peaks at lower frequencies ($F_\beta$ and $F_\gamma$) can be tracked up to 40$^{\circ}$ while $F_\zeta$, $F_\kappa$ and $F_\mu$, which are less intense, disappear at around 20$^\circ$. The angular dependence of these frequencies are displayed in Figs.~\ref{Ang_dep}(d) and \ref{Ang_dep}(e) and are well-described by the formula $F(\theta)=F(0^{\circ})/ \cos(\theta)$, consistent with the two-dimensional feature of Fermi surfaces. Thus, the angular dependence results in \KVSb\ are similar to the results in both \RbVSb\ and \CsVSb~\cite{Zhang2022, Wang2022a, Fu2021, Ortiz2021} and consistent with the two-dimensional nature of Fermi surfaces obtained by density functional theory (DFT) calculations of \KVSb~\cite{Tan2021,Uykur2021}.

 \begin{figure*}[!t]\centering
      \resizebox{16cm}{!}{
             \includegraphics{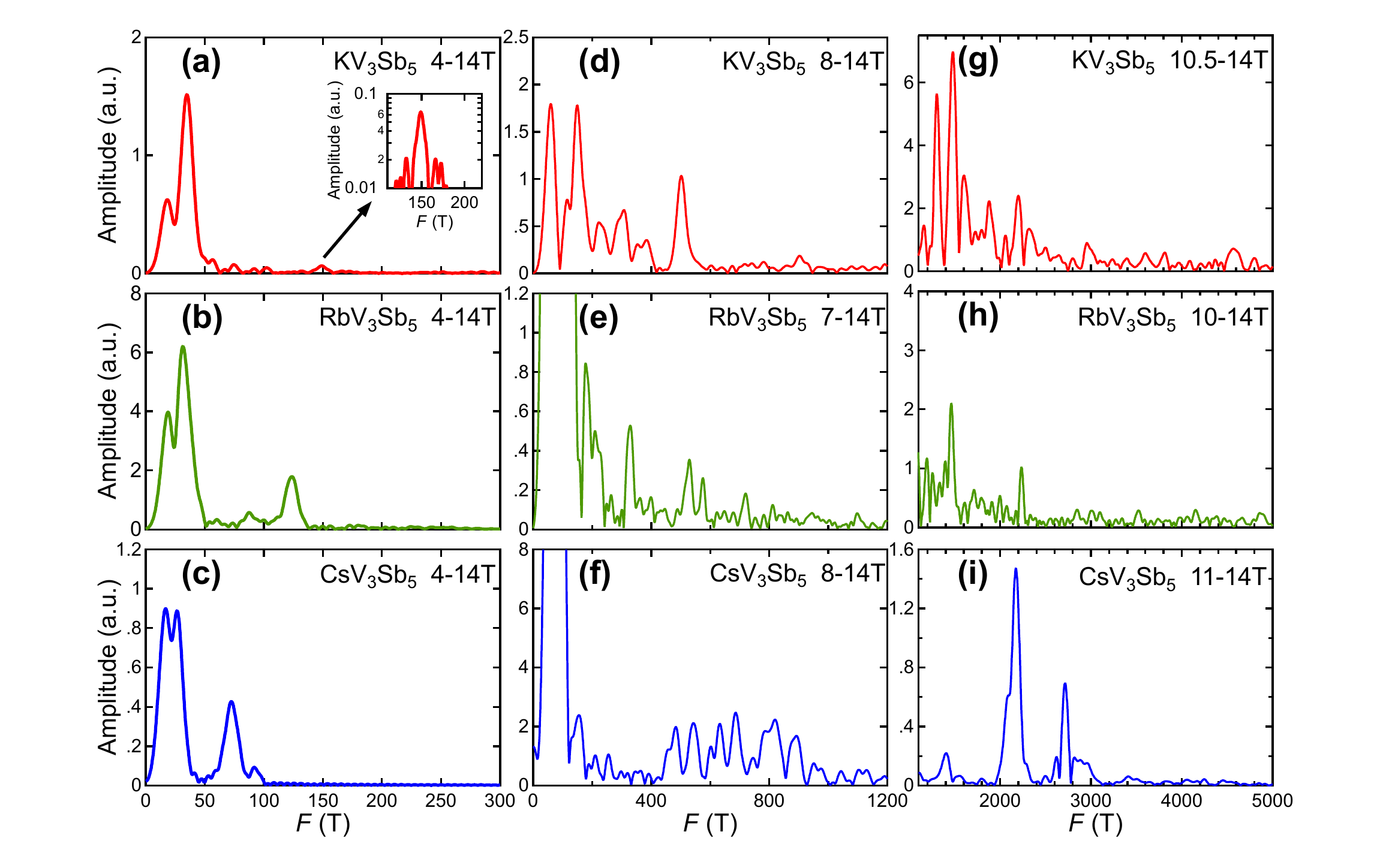}}      
             \caption{\label{AVSb} 
             Comparison of SdH oscillation spectra of \KVSb, \RbVSb\ and \CsVSb. (a-c) The FFT spectra from 0~T to 300~T, (d-f) the FFT spectra from 0 to 1200~T and (g-i) the FFT spectra from 1100~T to 5000~T. Different field windows for performing FFT were used for different frequency ranges. The inset in (a) offers a better view of $F_\gamma$ in \KVSb. Data of \RbVSb\ and \CsVSb\ come from Ref.~\cite{Wang2022a} and Ref.~\cite{Zhang2022}, respectively.
             }

\end{figure*}



\section{Discussion} 
The complex SdH spectrum of \KVSb, which features QO frequencies spanning a large range, is reminiscent of the spectra observed in the sister compounds \RbVSb\ and \CsVSb. Given that some of us have recently studied QOs in both \RbVSb~\cite{Wang2022a} and \CsVSb~\cite{Zhang2022} with the same experimental protocol, it is meaningful to compare our data across all three kagome metals \AVSb. Figure~\ref{AVSb} summarizes the SdH spectra of \KVSb, \RbVSb\ and \CsVSb\ at 2~K and with $B\,\parallelsum \,c$. To display the spectra in different frequency regions, the field window for performing the FFT has been carefully chosen and applied consistently across three systems.

Figures~\ref{AVSb}(a), \ref{AVSb}(b) and \ref{AVSb}(c) respectively show the low-frequency spectra of \KVSb, \RbVSb\ and \CsVSb, with the field range of $4-14$~T. Similar double-peak structure located below 50~T can be seen, in addition to another peak whose frequency decreases systematically when the atomic size of the alkali element increases: in \KVSb, \RbVSb\ and \CsVSb, the frequency is 149~T, 124~T and 72~T, respectively. According to previous studies~\cite{Fu2021}, the 72~T peak in \CsVSb\ corresponds to a small oval-shaped pocket near the L point in the first Brillouin zone. In \KVSb, a similar oval-shaped pocket was also reported near the L point in DFT calculations~\cite{Luo2022} and this could be assigned to the 149~T peak we detected.

To have a better view of the higher frequencies, narrower field windows have been used, resulting in the spectra displayed in Figs.~\ref{AVSb}(d)$-$\ref{AVSb}(i). Although the signal-to-noise ratios are different, similar features can be spotted: oscillation frequencies near 2200~T and around 1400~T, as shown in Figs.~\ref{AVSb}(g), \ref{AVSb}(h) and \ref{AVSb}(i). We note that the frequencies over 2000~T in Fig.~\ref{AVSb}(i) are beyond the expectation in DFT calculations taking into account a 2$\times$2$\times$4 distortion of \CsVSb~\cite{Zhang2022}, and their presence has been linked to orbital-selective charge doping. Meanwhile, the QO spectra of \KVSb\ and \RbVSb\ also contain frequencies beyond 2000~T, although they are less intense than the \CsVSb\ counterpart. Hence, further studies are needed to settle the origin of these high-frequency peaks. Finally, prominent differences appear in the mid-frequency range, see Figs.~\ref{AVSb}(d)-\ref{AVSb}(f). Contrary to \CsVSb, in which multiple frequencies with nearly equal spacing appear between 400 T to 1000 T, both \KVSb\ and \RbVSb\ show simpler spectra with better-defined, relatively more isolated peaks. 

At the qualitative level, the QO spectrum of \KVSb\ resembles the spectrum of \RbVSb\ more than that of \CsVSb. This could reflect the differences in the CDW distortion. In \KVSb\ and \RbVSb, a 2$\times$2$\times$2 distortion has been reported~\cite{Kang2022a}, while \CsVSb\ adopts a more complicated 2$\times$2$\times$4 distortion~\cite{Ortiz2021,Kang2022a} - the mid-frequency and the high-frequency spectra could be more sensitive to the precise nature of the CDW distortion. Nevertheless, all three kagome metals display rich QO spectra with recognizable similarity, quasi-2D Fermi surfaces as well as light effective masses.

\section{Conclusions}
In summary, we have conducted the SdH quantum oscillation measurements in high-quality single crystals of \KVSb. We resolve a large number of new frequencies up to 2202~T. The Lifshitz-Kosevich analysis reveals relatively small cyclotron effective masses and the angular dependence of the QO frequencies are consistent with two-dimensional Fermi surfaces. We further compare the QO spectra for all three kagome metals \AVSb. Our results show that the QO spectra in all three compounds are complicated because of the complex CDW order, and that large Fermi surfaces exist in these metals.  Our results fill in the gap of the QO study in \KVSb\ and provide valuable data to understand the band structure of all three members of \AVSb.
\\
\\

\begin{acknowledgments}
We acknowledge support by Research Grants Council of Hong Kong (CUHK 14301020, CUHK 14300722, A-CUHK402/19), CUHK Direct Grant (4053461, 4053408, 4053528, 4053463), the National Natural Science Foundation of China (12104384, 12174175) and the Shenzhen Basic Research Fund (JCYJ20190809173213150).\\ 
$^\S$W.Z. and Z.W. contributed equally to this work.
\end{acknowledgments}


\providecommand{\noopsort}[1]{}\providecommand{\singleletter}[1]{#1}%

\end{document}